\documentclass[12pt]{article}

\usepackage{epsf}

\def\fun#1#2{\lower3.6pt\vbox{\baselineskip0pt\lineskip.9pt
\ialign{$\mathsurround=0pt#1\hfil##\hfil$\crcr#2\crcr\sim\crcr}}}

\title{\bf Quadrupolar contact terms and Hyperfine Structure}
\author{ G.Karl and V. A. Novikov\footnote{On leave from ITEP, Moscow, Russia}
\\ Department of Physics, University of Guelph, Guelph, Canada,\\ Perimeter Institute, Waterloo, Canada}

\date{}
\begin{document}
\maketitle

\begin{abstract}
In the interaction of two electric quadrupoles, there is at short range a contact term proportional to the second derivative of a delta function. This contact term contributes to the hyperfine splitting of bound states of two particles with
 spin one or higher - for example the bound state of  $\Omega^{-}$ and a nucleus with spin one. The contact hyperfine splitting
occurs in states with orbital angular momentum unity ($p$-wave), in contrast to the Fermi contact interaction which is in $S$-states. We find that these
contact splittings will be observable with  $\Omega^{-}$ atoms and help measure the
quadrupole moment and charge radius of the hyperon.

\end{abstract}
\newpage

\section {\bf Introduction}

It is well known in the context of atomic hyperfine interactions that the magnetic dipole-dipole interaction has a contact term, called the Fermi
contact interaction \cite {Fermi}. A well known example is the splitting of $F=0$ from the $F=1$ states   of the $H$ atom, responsible for the microwave radiation of interstellar Hydrogen. It is  natural
to ask whether such contact interactions occur in higher multipoles, for example in the interaction of two electric quadrupoles. Only particles
with angular momentum $J=1$ or higher can have quadrupoles. We consider
here as an example the bound state of an $\Omega^{-}$ hyperon  ( J=3/2 ) to a nucleus of angular momentum one (say $Nitrogen 14$), though there are other possibilities like a bound state of an anti-deuteron to a nucleus, etc.
We find in addition to the well known long-range electrostatic
interaction of two quadrupoles ($\sim 1/r^5$) there are short range contributions some of which are proportional to the second derivative of
a delta function such as $\Delta\delta(\bf{r})$. These contact terms give rise to splitting in the $P$- wave ($L=1$) states of the bound system of an  $\Omega^{-}$
with a nucleus, or other systems with two quadrupoles. As far as we know this higher contact interaction has never been described, although it is a consequence of ordinary electromagnetism and quantum mechanics. We
find that this splitting may be useful to determine the electric quadrupole moment of the  $\Omega^{-}$, which is not yet known experimentally.

 In an ordinary atom, an electron is bound to the nucleus. The electron has spin, and associated with the spin there is a magnetic moment. This magnetic moment interacts with the magnetic fields in the atom and leads to splittings in the energy levels which are called fine structure- if
the nucleus itself has no magnetic moment, for example a spinless nucleus such as Helium four. However if the nucleus itself has spin and
magnetic moment, the nuclear magnetic moment (which is much smaller than
the electronic magnetic moment) gives rise to further splittings which
are called hyperfine structure. Other contributions to atomic hyperfine
structure arise from nuclear electric quadrupole moment ( which is
sensitive to electric field gradients at the nucleus). Higher nuclear
moments tend to produce smaller effects.
 In an $\Omega^{-}$ atom where the $\Omega^{-}$ is bound to a nucleus,
there are interesting effects coming from the magnetic moment of the
$\Omega^{-}$ (which is known experimentally) and the electric quadrupole
moment
(which is not known experimentally but estimated theoretically).
The fine structure in an $\Omega^{-}$ bound to a spinless nucleus was discussed by Goldhaber and Sternheimer \cite{Gold}. In the scenario of Ref. \cite{Gold} the electric quadrupole dominates over magnetic contribution to the fine structure because the electric quadrupole is assumed to be large. However later estimates of the $\Omega^{-}$ quadrupole, using $QCD$ inspired quark models \cite {Gershtein} are about a hundred times smaller than assumed in reference \cite {Gold}. This leads to a completely different scenario, in which the fine structure is dominated by magnetic interactions, and the effect of the electric quadrupole is to be found in hyperfine interactions. This is the scenario we discuss here by taking a nucleus which has an electric quadrupole and a (small) magnetic dipole to bind the
$\Omega^{-}$. The hyperfine effects will now be smaller than fine structure but still large enough to be observed. As far as we know these hyperfine effects have not been discussed in the literature. The most interesting of these  effects is in the interaction of two electric quadrupoles, where there is a contact term in the $l=1$ ($p$-wave) states. This contact term is discussed in Section 2 and Appendix 1, 2 . Then in Sections 3, 4 we discuss an application to $\Omega^{-}$ atoms. We conclude in Section 5 with a summary of the results.
\vspace{5mm}

\section {\bf Contact interaction of quadrupoles. }

\vspace{5mm}

The long range interaction of two quadrupoles $Q^{(1)}$ and $Q^{(2)}$ may be written
in terms of Cartesian components in the form \cite{quadr}:
$$
H_{tensor} = \frac{1}{48\pi} Q^{(1)}_{ij} \nabla_{i}\nabla_{j} (\frac {x_{k}x_{l}}{r^5}Q^{(2)}_{kl}) =
$$
$$ =\frac{1}{48\pi r^5} [2Q^{(1)}_{ij} Q^{(2)}_{ij}- 20Q^{(1)}_{ij}Q^{(2)}_{ik}n_{j}n_{k} + 35 Q^{(1)}_{ij}Q^{(2)}_{kl}n_{i}n_{j}n_{k}n_{l}]
\eqno (2.1)
$$
In equation (2.1) repeated indices are summed; in the first term the two
quadrupoles are fully contracted and in the next terms $n_{i} = x_{i}/r$, etc.
The interaction of two quadrupoles decreases very rapidly with distance-
like $r^{-5}$, as follows from dimensions- much faster than the fine structure coupling, which decreases only like $r^{-3}$. As a result the
quadrupolar coupling is more important for smaller orbits, at low quantum numbers. Therefore we will focus on low lying $P$-states of the atom. It can be checked that the tensor quadrupolar interaction (eq. 2.1) is ambiguous in $P$-states of the atom; the angular averaging vanishes
because the angular dependence of the interaction (when expressed in spherical harmonics) involves $Y_{4,m}(\theta,\phi)$, where the angles $\theta,\phi$ are those
of the vector connecting the two quadrupoles \cite{quadr}.The expectation value of a spherical harmonic $Y_{4,m}$ vanishes in a $P$-wave state. The radial integral on the other hand diverges. This ambiguity is similar to the
case of the dipole-dipole interaction in $S$-waves, where the contact interaction solves the ambiguity. Similarly here, instead of the tensor
interaction of two quadrupoles we have a contact term, involving the
derivative of a delta function, which generalizes the contact
interaction where there is only a delta function. The quadrupolar
contact coupling is a scalar interaction coupling the two quadrupole
tensors and second derivatives of a delta function (as required by
dimensional considerations) to a scalar. The precise formula for the
quadrupolar contact term is given below
$$
H_{cont}= -\frac {1}{36}[\frac{4}{7}(\nabla_j \nabla_k-\frac{1}{3} \nabla^2\delta_{jk})\delta^{(3)}({\bf {r}}) +\frac{2}{15 }\delta_{jk} \nabla^2\delta^{(3)}({\bf {r}})]Q_{ij}^{(1)}Q_{ik}^{(2)}.
\eqno(2.2)
$$

Equation (2.2) is derived in the  Appendix 1. In this formula $Q^{(1),(2)}$ are  the
quadrupole moments of the hyperon and the nucleus, and $ \nabla^2= \nabla_{i}\nabla_{i}$ is the Laplace operator. This contact term gives a finite contribution in a $P$-wave
state where the two spatial derivatives are taken up by the wave
function squared.

\vspace{5mm}

\section {\bf  Structure of $\Omega$ atom}

\vspace{5mm}
 Consider an $\Omega^{-}$ interacting with nucleus with charge $Ze$. In the first approximation this is Coulomb system with spectrum

$$
E_{n,l}= - \frac { Z^2\alpha^2}{2n^2}m,
\eqno (3.1)
$$
where $n,l$ are principal and angular momentum  quantum numbers, $m$ is the reduced mass
$$
m=\frac {m_{\Omega}}{(1+m_{\Omega}/M)}
$$
 $m_{\Omega}$ is $\Omega$ mass and $M$ is the mass of the nucleus. The corresponding Rydberg constant for heavy nucleus is
$$
R_{\Omega}= R_e \frac{m_{\Omega}}{m_e}= 44 keV.
\eqno (3.2)
$$

 The interaction of charge with magnetic  and quadrupole moments and of the multipole moments with each other is smaller than Coulomb interaction  of charges and can be considered as perturbation to Coulomb levels eq.(3.1).
In the case of ordinary atoms  perturbations split Coulomb levels into two main structures - fine structure (due to interaction of magnetic moment of electron with angular momentum of atom) and "then" fine structure levels into hyper-fine structure ( due to interaction of electron magnetic
moment with nuclear magnetic moment). In the case of ${\Omega^{-}}$ atom the splitting of the Coulomb levels can be more complicated because ${\Omega^{-}}$ has spin $S=3/2$
and magnetic and quadrupole interactions can have rather different scale of strength.
It is not a big deal to write equations for splitting for any possible case. In this paper we restrict ourselves to the case  most similar to the ordinary atoms, i.e.
we suppose that there is fine structure due to interaction
of $\underline {magnetic}$ moment of ${\Omega^{-}}$ with orbital angular momentum $\bf {L}$. That means in  particular that magnetic moment of nucleus is much smaller than the  magnetic moment of ${\Omega^{-}}$ and that corresponding interaction of quadrupole moments
with $\bf {L}$ is also much smaller.

For fine structure levels we have one "good" quantum number-total angular momentum $\bf {J}$ that is the sum of orbital angular momentum $\bf {L}$ and spin ${\bf S}$  of   ${\Omega^{-}}$:
$$
{\bf J} = {\bf L}+ {\bf S}.
$$

Each Coulomb level $ E_{n, l}$  splits in general into four levels  with different total angular momentum $j=l+3/2, j=l+1/2,  j=l-1/2, j=l-3/2$. The magnetic interaction for  ${\Omega^{-}}$ atoms is similar to the case of ordinary atoms
$$
V_{mag} = -\frac{Z\alpha}{2 m^{2}r^{3}}(g-1)\bf{L\cdot S},
\eqno (3.3)
$$
where factor $g$ relates magnetic moment with spin  {\boldmath {$\mu$}}$_{\Omega}= g{\bf {S}}$.

The average value of $1/r^3$ has to be calculated with unperturbed Coulomb wave functions
$$
<\frac{1}{r^3}>= \frac {(Z\alpha m)^3}{n^3 l (l+1/2)(l+1)}.
\eqno (3.4)
$$

 Thus the fine splitting is equal to

$$
\Delta E (n,l,j)= \frac{(Z\alpha)^{4}}{n^3}m(g-1) \frac{j(j+1)-l(l+1)-15/4}{2l(2l+1)(l+1)}.
\eqno (3.5)
$$
 Magnetic moment of  ${\Omega^{-}}$  is known experimentally $\mu_{\Omega}=-2.02 \pm0.05\;\mu_N$
\cite {PDG}.
Thus substituting these factors into $V_{mag}$ we get for fine splitting of $p$-wave levels into three states with
total angular moment $j=5/2, 3/2, 1/2:$
$$
\Delta E (n,l,j)= \frac{Z^4}{12 n^3}[j(j+1)-\frac{23}{4}]\;
(6.64 \;ev).
\eqno (3.6)
$$

This formula gives the scale and exact form of the fine splitting.

We are interested in the hyperfine splitting of these levels by interaction of $\Omega$  quadrupole moment  with nucleus. In this case only total angular moment of the system $\bf {F}=\bf{J}+\bf{I}$ is a good quantum number. (Here $\bf {I}$ is a nuclear spin).  The splitting due to long range interaction
was considered in ref. \cite{Gold}. But interaction of quadrupole moment with charged system has also local terms proportional to $\delta$-function and to it derivatives. Thus there should be additional  hyperfine structure for states with small angular momentum, i.e.  additional small shifts of s- and p- states. We are interested in hyperfine structure due to these contact terms.

 There are well-known local  terms, such as Fermi contact interaction of two magnetic moments {\boldmath{$\mu$}}$_{1,2}$

$$
U(x)=- \frac{2}{3}( \mu_i^{(1)} \mu_i^{(2)})\delta^{(3)}({\bf r}).
\eqno (3.7)
$$
Fermi term gives contribution into $s$-wave only and it does not compete with hyper-fine structure due to quadrupoles in $p$-waves.

Interaction of charge $e_{1,2}$ with e.m. radius $r_{2,1}$ and radius-radius interaction are also local
$$
U(x)= - e_1 e_2[ \frac{1}{6}(r_1^2+r_2^2)\delta^{(3)}({\bf r}) + \frac {1}{36} r_1^2r_2^2 \nabla^2 \delta^{(3)}({\bf r})].
\eqno(3.8)
$$
First term in eq.(3.8) contributes into $s$-wave states shift, the second one into $s$- and $p$-wave states. But both of this terms do not depend on the spin. Thus they contribute into shift but not into splitting of states with different $J$ and/or $F$.

Contact terms from eqs. (3.7) and (3.8) are well known and their derivation can be found in text-books on Classical Electrodynamics ( see, e.g., ref. \cite {Jack}).
Similar local terms exist also in the interaction of quadrupole moments, which we cannot find in the literature.

There are two such terms ( for derivation see Appendix 1):

Interaction of quadrupole moment $Q_{ij}$ with e.m. radius $r$

$$
V(x)= -\frac {1}{36 } [e_{1}r_{1}^{2} Q_{ij}^{(2)}+ e_{2} r_{2}^{2} Q_{ij}^{(1)}]\nabla_i \nabla_j\delta^{(3)}({\bf r}),
\eqno (3.9)
$$

 and quadrupole-quadrupole interaction

$$
V(x)= -\frac {1}{63}Q_{ij}^{(1)}Q_{ik}^{(2)}[\nabla_j \nabla_k-\frac{1}{10} \nabla^2\delta_{jk}]\delta^{(3)}({\bf {r}}) .
\eqno(3.10)
$$

These contact terms contribute to splitting of $s$- and $p$-waves states.

Notice that {\bf long-range} interaction of two quadrupoles does not contribute into splitting of $s$- and $p$-states. Any matrix element of this $d$-wave interaction vanishes identically for $s$- and $p$-states.

We are interested in determination of the value of $\Omega$ quadrupole moment from the spectrum of $\Omega$ atom.  Thus we concentrate on hyperfine structure of p-states.

There are three sources of splitting for $p$-states as it is seen from eqs. (3.9), (3.10).

First there is $\Omega$ quadrupole - nuclear radius interaction.
It looks like

$$
V(x)= \frac {Ze}{36 } r_{(Z)}^{2} Q_{ij}^{(\Omega)}\nabla_i \nabla_j\delta^{(3)}({\bf r}).
\eqno (3.11)
$$

Quadrupole moment of $\Omega$ is related to the spin operator ${\bf S}$  of $\Omega$
$$
Q^{\Omega}_{ij}=\frac {3Q^{(\Omega)}}{2S(2S-1)}(S_i S_j + S_jS_i - \frac{2}{3}S(S+1) \delta_{ij})= \frac {Q^{(\Omega)}}{2}(S_i S_j + S_jS_i - \frac{5}{2}\delta_{ij}).
\eqno (3.12)
$$

Thus contact term eq.(3.11) does not depend on the value of spin $\bf{I}$ of nucleus. As a result the 
corresponding shift of $p$-level can depend only on the value of $\bf{J}= \bf{L} + \bf{S}$, i.e. this contact 
term contributes into fine splitting of Coulomb $p$-levels, not into hyper-fine ones. According to our 
assumption fine structure is mainly determined by interaction of $\Omega$ magnetic moment with orbit. 
Thus interaction eq.(3.11) slightly change this fine shift of levels with different angular momentum $j$ but does not  split them.

There is a similar term in interaction of nuclear quadrupole moment $Q_{ik}^{(Z)}$ with e.m. radius of $\Omega$

$$
V({\bf r})= \frac {(-e)}{36 } r_{(\Omega)}^{2} Q_{ij}^{(Z)}\nabla_i \nabla_j\delta^{(3)}({\bf r})
\eqno (3.13)
$$
with

$$
Q^{(Z)}_{ij}=\frac {3Q^{(Z)}}{2I(2I-1)}\{I_i I_j + I_jI_i - \frac{2}{3}I(I+1) \delta_{ij}\}.
\eqno (3.14)
$$
This interaction leads to hyperfine splitting of $p$-levels
indeed. The mathematics that helps to calculate matrix elements of tensor operators  between wave functions 
from multiplets with different $j$ and $f$ is very similar to the case of ordinary atomic physics \cite {LL1}. The 
result of this calculation is following
$$
\Delta E^{(1)} (n,l=1,j,f)= \frac{e}{6}Q^{(Z)}r^2_{(\Omega)}|f(0)|^2 X^{(1)}(j) T(f, j),
 \eqno(3.15)
$$
where $f(0)$ is the derivative of Coulomb $l=1$ wave function at origin
(see \cite{LL1})
$$
|f(0)|^2= \frac{n^2 -1}{3\pi n^5} (m Z\alpha)^5,
\eqno (3.16)
$$
 and $T(f,j)$ and $X(j)$ are spin-factors
$$
T(f,j)= I_i I_j\{J_i J_j + J_jJ_i - \frac{2}{3}j(j+1) \delta_{ij}\}= 2({\bf J\cdot I})^2 +({\bf J\cdot I}) - \frac{2}{3} I(I+1)j(j+1),
\eqno (3.17)
$$
and
$$
X^{(1)}(j)=\frac{3[j(j+1)-77/12]}{8j(j+1)}.
\eqno (3.18)
$$
Here
$$
 J_i I_i  ={\bf J\cdot I} =\frac{1}{2} [f(f+1)-I(I+1)-j(j+1)].
$$

For $I=1$ nucleus and for $p$-states the only interesting  states are those with $j=3/2$ and  $j=5/2$. 
For this cases $X^{(1)}(j=3/2)=-4/15$ and   $X^{(1)}(j=5/2)= 1/10$. For $j=1/2$, $T(f,j)$ vanishes, as it 
should, since one can't have a quadrupole at $j=1/2$.

 In this paper we are interested mainly in  {\bf $\Omega$ quadrupole - nuclear quadrupole} contact interaction

$$
V(x)= -\frac {1}{63}Q_{ij}^{(\Omega)}Q_{ik}^{(Z)}[ \nabla_j \nabla_k-\frac{1}{10} \nabla^2\delta_{jk}]\delta({\bf{r}}).
\eqno(3.19)
$$
Matrix elements of tensor operators for this interactions can be calculated in the similar way. The result is
$$
\Delta^{(2)} E (n,l=1,j,f)= -\frac{1}{21}Q^{(Z)}Q^{(\Omega)}|f(0)|^2 X^{(2)}(j) T(f, j),
\eqno(3.20)
$$
with spin factor
$$
 X^{(2)}(j)= -\frac{1}{16}\{j(j+1)-\frac{69}{10} -\frac {49\cdot 31}{80 j(j+1)}\}.
\eqno (3.21)
$$
For $I=1$ nucleus and for $p$-states the only interest is in $j=3/2$ and  $j=5/2$, as before. 
For this cases $X^{(2)}(j=3/2)=77/150$ and   $X^{(2)}(j=5/2)=1/50$.

 For the derivation of eqs. (3.18) and (3.21) see next section.

\section{Calculation of spin-factors}

The calculation of matrix elements of tensor operators between states that correspond to irreducible 
representation of rotation group is well-known (see, e.g \cite{LL1}). In this section we present some 
intermediate  formulae that help to get eqs. (3.18), (3.21).

Matrix elements of any tensor in the subspace of multiplets with given value of {\bf{J}}  can be represented 
as a matrix elements of appropriate combination of operators {\bf\^J}.

We start with the state with given ${\bf{L}}$, namely with subspace of $p$-states. In the special basis where $l=1$ 
wave function is represented as a vector-function ${\bf\Psi}(x) = {\bf x} f(r)\;\; ( r=|{\bf x}|)$ one can check that

$$ \nabla_i \nabla_j\delta^{(3)}({\bf r})= |f(0)|^2 \{\hat{L}^2\delta_{ij}-\hat{L}_{i}\hat{L}_{j}-\hat{L}_{j}\hat{L}_{i}\}
$$

Thus interaction of nuclear quadrupole moment $Q_{ik}^{(Z)}$ with e.m. radius of $\Omega$ can be rewritten as

$$
V(x)= \frac {e r_{(\Omega)}^{2}}{36 } Q_{ij}^{(Z)}|f(0)|^2 \{\hat{L}_{i}\hat{L}_{j}+\hat{L}_{j}\hat{L}_{i}\}.
\eqno (4.1)
$$
( Term $\sim \delta_{ij}$ does not contribute into interaction since tensor $Q_{ij}^{(Z)}$ is traceless).

Second step is averaging of this operator over states with given {\bf J}. At this step any tensor can be 
constructed from operators $\hat{J}_i$ so that
$$
\hat{L}_{i}\hat{L}_{j}+\hat{L}_{j}\hat{L}_{i}-2/3\hat{L}^2\delta_{ij} = X^{(1)}(j)\cdot \{\hat{J}_{i}\hat{J}_{j}+\hat{J}_{j}\hat{J}_{i}-2/3\hat{J}^2\delta_{ij}\},
\eqno (4.2)
$$
 where $X^{(1)}(j)$ is an unknown constant. To calculate this constant we use the following relations

$$
T_0=\hat{J}_{i}\{\hat{J}_{i}\hat{J}_{j}+\hat{J}_{j}\hat{J}_{i}-\frac{2}{3}\hat{J}^2\delta_{ij}\}\hat{J}_{j} = 4/3 j(j+1)(j-1/2)(j+3/2);
\eqno (4.3)
$$
$$
T_1=\hat{J}_{i}\{\hat{L}_{i}\hat{L}_{j}+\hat{L}_{j}\hat{L}_{i}-\frac{2}{3}\hat{L}^2\delta_{ij}\}\hat{J}_{j} =
= 2({\bf J\cdot L})^2 -({\bf J\cdot L})-\frac{2}{3} l(l+1)j(j+1)=
$$
$$=\frac{1}{2} (j-1/2)(j+3/2)\{j(j+1)-77/12\},
\eqno (4.4)
$$
where in the last line we take $l=1$. Thus for constant $X^{(1)}(j)$ one gets relation
$$
X^{(1)}= T_1/T_0,
$$
that is equivalent to eq.(3.18) from the previous section.

Quadrupole-quadrupole interaction for the states with given {\bf L} can be rewritten as

$$
V(x)= -\frac {1}{63}|f(0)|^2 Q_{ij}^{(\Omega)}Q_{ik}^{(Z)}\{\frac{9}{10}\hat{L}^2\delta_{jk}-\hat{L}_{j}\hat{L}_{k}-\hat{L}_{k}\hat{L}_{j}\} ,
\eqno(4.5)
$$
where the quadrupole tensors are
$$
Q^{(Z)}_{ij}=\frac {3Q^{(Z)}}{2I(2I-1)}\{I_i I_j + I_jI_i - \frac{2}{3}I(I+1) \delta_{ij}\},
\eqno (4.6)
$$
and
$$
Q^{(\Omega)}_{ij}=\frac {3Q^{(\Omega)}}{2S(2S-1)}\{S_i S_j + S_jS_i - \frac{2}{3}S(S+1) \delta_{ij}\}.
\eqno (4.7)
$$
 For multiplet with given total angular moment {\bf J}

$$
V(x)= -\frac {1}{63}|f(0)|^2 Q_{ij}^{(Z)}<Q_{ik}^{(\Omega)}\{\frac{9}{10}\hat{L}^2\delta_{jk}-\hat{L}_{j}\hat{L}_{k}-\hat{L}_{k}\hat{L}_{j}\}>_{J} ,
\eqno(4.8)
$$
where $<
..>_{J}$  is average over multiplet with given {\bf J}
 and
$$
<Q_{ik}^{(\Omega)}\{\frac{9}{10}\hat{L}^2\delta_{jk}-\hat{L}_{j}\hat{L}_{k}-\hat{L}_{k}\hat{L}_{j}\}>_{J}=X^{(2)}(j) \{\hat{J}_{i}\hat{J}_{j}+\hat{J}_{j}\hat{J}_{i}-\frac{2}{3}\hat{J}^2\delta_{ij}\}.
\eqno (4.9)
$$

To calculate $X^{(2)}(j)$ we need new relations
$$
T_2=\hat{J}_{i}\{\hat{S}_{i}\hat{S}_{j}+\hat{S}_{j}\hat{S}_{i}-\frac{2}{3}\hat{S}^2\delta_{ij}\}\hat{J}_{j} = \frac{1}{2}(j-1/2)(j+3/2)\{j(j+1)- 7/4\},
\eqno (4.10)
$$
$$
T_3=\hat{J}_{i}\{\hat{S}_{i}\hat{L}_{j}+\hat{S}_{j}\hat{L}_{i}-\frac{2}{3}{\bf{S\cdot L}}\delta_{ij}\}\hat{J}_{j} =
$$
$$=\frac{1}{6} (j-1/2)(j+3/2)\{j(j+1)+49/4\}.
\eqno (4.11)
$$
After some algebra one gets eq.(3.21) for $X^{(2)}(j)$.

\section{Numerical estimates}

 Thus for hyperfine splitting we get
$$
\Delta E (n,l=1,j,f)= \frac{1}{6}Q^{(Z)}|f(0)|^2 T(f, j)\{ X^{(1)}(j) e r^{2}_{\Omega} -\frac{2}{7}X^{(2)}(j)Q^{(\Omega)}\} ,
\eqno(5.1)
$$
 with
$$
|f(0)|^2= \frac{n^2 -1}{3\pi n^5} (m Z\alpha)^5,
\eqno (5.2)
$$
$$
T(f=1/2)=5, T(f=3/2)=-4, T(f=5/2)=1
\eqno (5.3)
$$
for $j=3/2$ and
$$T(f=3/2)= 28/3, T(f =5/2)=-32/3, T(f=7/2)=10/3
\eqno (5.4)
$$
for $j=5/2$.
One can check that  hyperfine splitting satisfies well-known Sum Rules
$$
2 T(1/2) + 4 T(3/2)+ 6 T(5/2) =0
\eqno (5.5),
$$
and $$
4 T(3/2)+ 6 T(5/2) +8 T(7/2)=0.
\eqno (5.6)
$$

Unknown quantities in eq.(5.1) are electric radius  $e r^{2}_{\Omega}$ and quadrupole moments $Q^{(Z)}$ and $Q^{(\Omega)}$.
If we take  $e r^2_ {\Omega} \sim 10^{-2} fermi^2$ and $Q^{(\Omega)}\sim 10^{-2} fermi^2$ we get that for $Z=7$ hyperfine splitting is of the order of $keV$. We believe that such splitting can be measured experimentally.

 The main background that interferes with hyperfine splitting is strong interaction of $\Omega$ with nucleus \footnote{ We are grateful to M.Pospelov who  brought our attention to this subject.}.
Little is known about strong interaction of $\Omega$. To get an estimate for the shift and  width of Coulomb levels in $\Omega$ atoms due to strong interaction we looked at $p$-levels of antiprotonic atoms \cite {Klempt}. According to these data for antiprotonic hydrogen atom $\Gamma(2p) \sim 10^{-1} eV$ and for antiprotonic deuterium atoms $\Gamma(2p) \sim 5\cdot 10^{-1} eV$.
For heavier atoms widths of the states grows rapidly. Thus
 $\Gamma(2p) \sim 660 eV$ for $6Li$ \footnote { We are grateful to L.Bogdanova, B.Kerbikov, A.Kudryavtsev, D.Gotta and D.C. Bailey for discussion of the current experimental results}.

 We suppose  that $\Omega$ interacts with nucleus less than antiproton. Thus there is a good chance that strong interaction does not screen hyperfine structure of
$\Omega$ atoms due to contact interactions. It is a challenge for experimentalist to measure quadrupole moment of $\Omega$.

\section {Summary and Conclusions}

We have discussed the hyperfine spectrum of a bound system of $\Omega^{-}$ and a nucleus 
of angular momentum one, in a scenario which corresponds to a small electric quadrupole moment 
for the   $\Omega^{-}$, of the order  $10^{-2}$ fermi square. We find that there is a quadrupolar
contact term which contributes to the \underline{orbital} angular momentum one states. Our 
discussion suggests that contact terms are a universal property of interaction between multipoles, 
and not restricted only  to the interaction of two dipoles- electric or magnetic. However, multipoles 
higher than quadrupoles require two particles of spin $3/2$, and the number of such systems is 
quite small. We assume that the nucleus which binds the $\Omega^{-}$ has only an electric quadrupole
moment and no magnetic dipole, an oversimplified case, which is approximated by the nucleus 
$Nitrogen 14$. We do not imply that this is the optimal case to determine the quadrupole moment of the
$\Omega^{-}$. It remains to be found by experiment if the scenario we  discuss
 is in fact useful.

\section {Acknowledgments}
This research was supported by the NSERC and partly by RFBR grant
05-02-17203. One of the author (V.N.) is indebted to Physics
Department in University of Guelph and to Perimeter Institute for
hospitality.

\newpage

{\bf \Large Appendix 1. }

\bigskip

{\bf \Large Electromagnetic interaction of two particles}

\bigskip

 The interaction of two  multipole moments at long distances is well known and can be found in text-books on classical electrodynamics. In this paper we are mainly interested  in the contact interactions of the moments. These terms are less familiar. We derive them using a Feynman diagram technique in non-relativistic notations.

Consider the scattering of two particles with charges $( e_1, e_2 )$,  dipole moments $({\bf d}_1, {\bf d}_2)$, etc.
 The scattering amplitude $T(q)$ is given by one photon exchange diagram  and is equal to
$$
T(q) = - \frac {1}{q^2} [ \rho^{(1)}(q)\rho^{(2)}(-q) - {\bf j}^{(1)}(q){\bf j}^{(2)}(-q)],
\eqno(A1.1)
$$
where  $\rho(q)$ and ${\bf j}(q)$ are the matrix elements of the density and of the current of the particle.
In the non-relativistic approximation $T(q)=-U(q)$, where $U(q)$ is the Fourier transform of the potential energy. This gives  a systematic way to calculate potential energy between two particles with different multipoles moments.

To perform the calculation we need first of all the expression for the matrix element of the current operator
$j(q)=<p_2|\hat j|p_1>$ for a particle with electric dipole moment ${\bf d}$, with magnetic moment  {\boldmath ${\mu}$}, with quadrupole moment $Q_{ij}$, etc. The expression for matrix element of electromagnetic current $j_{\mu}= (\rho, \bf{j})$ is the following (for derivation see Appendix 2):

$$
 \rho(q)= j_0(q) \simeq e [ 1-\frac{1}{6} r^2 q^2] - i d_i q_i - \frac{1}{6} Q_{ij}q_i q_j + ... ,
$$
$$
 j_l (q) \simeq  \frac{e}{2}(v_1+v_2)_l-i{\it e}_{ikl}\mu_i q_k + ... ,
\eqno(A1.2)
$$
where ${\bf q}={\bf p}_2 - {\bf p}_1 $ is momentum transfer and ${\bf v} = {\bf p}/m$ is velocity
of a particle. We introduced form-factor $F(q^2)\simeq  1- (r^2 q^2)/6$ to take into account the distribution of charge. Here $r^2$ is the e.-m. radius of the distribution.
  The expression for the current density operator
${\hat{\bf{j}}}$ for a particle with charge $e$ and magnetic moment {\boldmath{$\mu$}} moving in a magnetic field is well
known and can be found in courses of quantum mechanics (see, e.g., Landau and Lifshitz \cite {LL1}). For the derivation of other terms in (A1.2) see Appendix 2.

Consider now how this procedure works in case of
{\bf charge-charge} interaction. The scattering amplitude is
$$
T(q) = - U(q)=-\frac{e_1 e_2}{q^2}[ 1 - \frac{1}{6}q^2(r_1^2+r_2^2)+\frac {1}{36} r_1^2r_2^2 q^4].
$$
In coordinate representation $1/q^2$ corresponds to $1/4\pi r$, constant in $q$-space corresponds to $\delta$-function in $x$-space, and $q^2$ corresponds to the second derivative of $\delta$- function:

$$
 ( 1/q^2)  \Longrightarrow \frac{1}{4\pi r};
$$

$$
  1  \Longrightarrow \delta^{(3)} ({\bf r});
$$

$$
 (q_i q_j)  \Longrightarrow -\nabla_i\nabla_j \delta^{(3)} ({\bf r}).
$$

 As a result $U(x)$ is given by
$$
U(x)= \frac{e_1 e_2}{4\pi r}- \frac{e_1 e_2}{6}(r_1^2+r_2^2)\delta^{(3)}({\bf r}) - \frac {e_1 e_2}{36} r_1^2r_2^2 \nabla^2 \delta^{(3)}({\bf r}).
\eqno(A1.3)
$$
We get long-range Coulomb term and \underline {contact} interactions of charge with charge radius and of two charge radii.

Consider {\bf dipole-dipole} interaction.
In momentum space
$$
U(q)=\frac {({\bf d}^{(1)}{\bf q})({\bf d}^{(2)}{\bf q})}{q^2}.
$$
The product $q_i q_k$ can be decomposed into two irreducible
tensors so that
$$
\frac{q_i q_k}{q^2}=(\frac{q_i q_k}{q^2}-\frac{1}{3} \delta_{ik}) + \frac{1}{3}  \delta_{ik}
\eqno(A 1.4)
$$

Irreducible tensors in momentum space  correspond to irreducible tensors in coordinate space.
Thus Fourier transform of $({q_i q_k}/{q^2}-{1}/{3} \delta_{ik})$ is traceless and symmetric second rank tensor in coordinate space made from the second derivatives of $1/(4\pi r)$, i.e.
$$
(\frac{q_i q_k}{q^2}-\frac{1}{3} \delta_{ik})\Longleftrightarrow
(\frac {-3 n_i n_k + \delta_{ik}}{4\pi r^3}),
$$
where ${\bf n}= {\bf r}/{r}$. Using irreducible tensors we automatically subtract contact terms $\sim \delta(x)$ from long-range potential. In spherical coordinates, the irreducible tensor of the second rank is proportional to the second spherical harmonic $Y_{2,m}(\theta,\phi)$.

The second term $\sim\delta_{ik}$ in eq (A 1.4) is constant in $q$- space. In coordinate space gives $\delta$-function, i.e.
$$
  1  \Longleftrightarrow \delta^{(3)}({\bf r})
$$
In this way we get well known long-range potential for two electric dipoles in p-wave
and point-like interaction of electric dipoles in s-wave

$$
U(x)=\frac {({\bf d}_1 {\bf d}_2) - 3({\bf d}_1{\bf n})({\bf d}_2{\bf n})}{4\pi r^3}+ \frac{1}{3}({\bf d}_1{\bf d}_2) \delta^{(3)}({\bf r})
\eqno (A1.5)
$$

{\bf Magnetic moment - magnetic moment} interaction  is very similar to electric moments interaction. The only difference is that {\boldmath {$\mu$}} contribute into {\bf j} ( not to $\rho$)
and that {\boldmath{$\mu$}} is axial vector so that instead of dot-products we have cross-products of vectors:

$$
U(q)= -\frac {1}{q^2}(e_{ikl}q_{k}\mu_l^{(1))})(e_{imn}q_{m}\mu_n^{(2)})
\eqno (A1.6)
$$
Transforming this expression to coordinate space we find

$$
U(x)=\frac {\mu_i^{(1)} \mu_i^{(2)} - 3(\mu_i^{(1)}n_i) (\mu_k^{(2)}n_k)}{4\pi r^3}- \frac{2}{3}( \mu_i^{(1)} \mu_i^{(2)})\delta^{(3)}({\bf r})
\eqno (A1.7)
$$
Long-range interaction is known from magneto-statics.
The result for contact term was found originally by Fermi and independently by Casimir in the framework of relativistic Dirac equation. Later it was derived using non-relativistic  Breit equation in the book \cite{Bethe}.
Elementary derivation of contact terms in magneto-statics is also known (see  \cite {Jack} ).

We are now ready to derive contact terms for quadrupole interaction.

{\bf Quadrupole-electric radius} interaction in momentum space
looks as
$$
V(q)= \frac {1}{36} [ e_{1} r_{1}^{2} Q_{ij}^{(2)}+ e_{2}r_{2}^{2}Q_{ij}^{(1)}] q_i q_j.
\eqno (A1.8)
$$
In coordinate space it gives

$$
V(x)= -\frac {1}{36 } [  e_{1} r_{1}^{2} Q_{ij}^{(2)}+  e_{2}r_{2}^{2} Q_{ij}^{(1)} ]\nabla_i \nabla_j\delta^{(3)}({\bf r}).
\eqno (A1.9)
$$

This interaction vanishes for s-wave but is nonzero for p-wave.

{\bf Quadrupole-quadrupole} interaction in momentum space has the following form
$$
V(q)= \frac {q_i q_j q_k q_l}{36 q^2} Q_{ij}^{(1)}Q_{kl}^{(2)},
\eqno (A1.10)
$$
where we have to express the product of the four momentum factors in terms of irreducible tensors. Different tensors
 represent interaction in $d$-, $p$- and $s$-wave.

 We construct irreducible tensor $T_{ijkl}^{(4)}$ as a sum of all possible symmetric terms:
$$
T_{ijkl}^{(4)}= q_i q_j q_k q_l + A q^{2}( \delta_{ij} q_kq_l +\delta_{kl} q_iq_j +\delta_{ik} q_jq_l +\delta_{jl} q_iq_k +\delta_{il} q_jq_k +\delta_{jk} q_iq_l)+
$$
$$
+ B q^{4} (\delta_{ij}\delta_{kl}+\delta_{ik}\delta_{jl}+\delta_{il}\delta_{jk}).
$$
Irreducible tensor $T_{ijkl}^{(4)}$ has to be  traceless.
The constraint $T_{iikl}^{(4)}=0$  gives $ A= -(1/7)$ and $B=(1/35)$.

Thus eq. (A 1.10) can be rewritten in following form
$$
V(q)= \frac{1}{36q^2}\{ Q_{ij}^{(1)}Q_{kl}^{(2)}T_{ijkl}^{(4)}\} + \frac{1}{63}\{ Q_{ij}^{(1)}Q_{ik}^{(2)} q_jq_k-\frac{q^2}{10} Q_{ij}^{(1)}Q_{ij}^{(2)}\}.
\eqno (A 1.11)
$$
In coordinate space  irreducible tensor $T_{ijkl}^{(4)}$ is
proportional to fourth harmonic $Y_{4,m}(\theta,\phi)$ . Thus this term in eq. (A 1.11) represents the well-known long-range interaction of two quadrupoles in d-wave (see eq. (2.1)).

When we go to coordinate space

$$
 ( 1/q^2)  \Longrightarrow \frac{1}{4\pi r}
$$

$$
 (q_i q_j)  \Longrightarrow -\nabla_i\nabla_j\delta^{(3)}(\bf r) .
$$
In this way differentiating $1/r$ we get the
standard formula for quadrupole-quadrupole interaction at long distance. Second term in eq. (A 1.11) is quadratic polynomial in momentum space. In coordinate space it corresponds to second derivative of delta function
$$
V_{cont}(x)= -\frac {1}{63}Q_{ij}^{(1)}Q_{ik}^{(2)}\{\nabla_j \nabla_k-\frac{1}{10} \nabla^2\delta_{jk}\}\delta ^{(3)}(\bf r) ,
\eqno(A1.12)
$$
which is equivalent to eq.(2.2) used in the text. This result is new.

\vspace{10mm}

{\bf \Large Appendix 2.}

\bigskip

{\bf\Large Electromagnetic vertex}

\bigskip

One can find the standard derivation for the current density operator  ${\hat{\bf{j}}}$ for a particle with charge $e$ and magnetic moment {\boldmath{$\mu$}} moving in a magnetic field in quantum mechanics text-books (see, e.g., Landau and Lifshitz \cite {LL1}).
The matrix element of this operator between two states with momenta ${\bf p}_1$ and  ${\bf p}_2$ is

$$
<p_2|{\hat j}_i| p_1> = \frac{e}{2} ({v}_1 +{v}_2)_i + i e_{ikl}q_k\mu_l ,
\eqno (A2.1)
$$
where ${\bf q}={\bf p}_2 - {\bf p}_1 $ is momentum transfer and ${\bf v} = {\bf p}/m$ is velocity
of a particle. For the charge density $\rho = j_0$ the matrix element is

$$
<p_2| \hat \rho| p_1> = e F(q^2)\simeq e ( 1-\frac{1}{6} r^2 q^2) .
\eqno (A2.2)
$$

We need similar expression for the electromagnetic current of a particle with electric dipole moment
${\bf d}$, with quadruple moment $Q_{ij}$ ( and with magnetic moment as well). One can derive this expression classically without 
any reference to quantum mechanics.

That is two-steps procedure (see also ref. \cite{Ber}).
First, notice that the interaction of the particle with electromagnetic field $A_{\mu}=( \phi, \bf{A})$ is described by currents
$$
H_{int} = \rho \phi - \bf{j} \bf{A}.
\eqno (A2.3)
$$
This is the
definition of the current.

Second, notice that in the static field the interaction energy of a particle is given by the multipole expansion. By definition of the multipole moments $\bf{d}$, {\boldmath${\mu}$}, $Q_{ij}$, etc, we have

$$
H_{int}\simeq e\phi(0) - d_i E_i -\frac{1}{6} Q_{ij}\frac{\partial E_j}{\partial x_i}-\mu_i H_i =
$$
$$
=[e + d_i\nabla_i +
\frac{1}{6} Q_{ij}{\nabla_{i}}{\nabla_{j}}]\phi(x)|_{x=0}-{\it e}_{ikl}{\mu}_{i}{\nabla}_{k}A_{l}(x)|_{x=0} ,
\eqno(A2.4)
$$
where $E_{i}(x)= -\nabla_{i}\phi(x)$,  $H_i = {\it e}_{ikl}{\nabla}_{k}A_{l}(x)$ are electric and magnetic fields respectively.

Comparing eqs.(A2.3) and (A2.4)
we find the contribution of electric dipole, magnetic dipole  and quadrupole moments into e.-m. current.

 In coordinate space it is equal to the sum of differential operators
$$
\hat j_0 = e + d_i \nabla_i + \frac{1}{6} Q_{ij}\nabla_i\nabla_j + ...,
$$
$$
\hat j_l = {\it e}_{ikl}\mu_i\nabla_k.
\eqno(A2.5)
$$
In momentum space
$$
j(q) = \int j(x) exp(-i{\bf q}{\bf x}) d^3x
$$
it can be written as
$$
 j_0 = e - i d_i q_i - \frac{1}{6} Q_{ij}q_i q_j + ... ,
$$
$$
 j_l = -i{\it e}_{ikl}\mu_i q_k + ... .
\eqno(A2.6)
$$

Fourier transform of the classical current corresponds to the matrix element of the quantum operator  $j(q) = <p_2|\hat j|p_1>$.

Thus we get the complete representation of the matrix element
of the current operator in momentum  space
$$
 j_0(q) \simeq e [ 1-\frac{1}{6} r^2 q^2] - i d_i q_i - \frac{1}{6} Q_{ij}q_i q_j + ... ,
$$
$$
 j_l (q) \simeq  \frac{e}{2}(v_1+v_2)_l-i{\it e}_{ikl}\mu_i q_k + ... .
\eqno(A2.7)
$$

In the language of Feynman diagrams eq.(A2.7) corresponds to
the vertex operator that describes interaction of the particle with photon.

\end{document}